\documentclass[referee,floatfix]{aa} 
\usepackage{graphicx}
\begin{document}
   \title{Observation of EGRET Gamma-Ray Sources by an Extensive Air Shower Experiment}


   \author{M. Khakian Ghomi
          \inst{} \fnmsep \thanks{\email{khakian@mehr.sharif.edu}}
          \and
           M. Bahmanabadi
          \inst{} \fnmsep \thanks{\email{bahmanabadi@sharif.edu}}
          \and
          J. Samimi
          \inst{} \fnmsep \thanks{\email{samimi@sharif.edu}}
          }

   \offprints{M. Khakian Ghomi}

   \institute{Department of Physics, Sharif university of
              Technology, P.O. Box 11365--9161, Tehran, Iran
         }

   \date{Received *** ; accepted ***}

\abstract{ Ultra-high-energy (E$>$100 Tev) Extensive Air Showers
(EASs) have been monitored for a period of five years (1997 -
2003), using a small array of scintillator detectors in Tehran,
Iran. The data have been analyzed to take in to account of the
dependence of source counts with respect to the zenith angle.
During a calendar year different sources come in the field of view
of the detector at varying zenith angles. Because of varying
thickness of the overlaying atmosphere, the shower count rate is
extremely dependent on zenith angle which have been carefully
analyzed over time. (\cite{magnetic}).High energy gamma-ray
sources from EGRET third catalogue where observed and the data
were analyzed using an excess method. A number of upper limits
for a number of EGRET sources were obtained, including 6 AGNs or
probably AGNs and 4 unidentified sources.
   }
   \authorrunning{M. Khakian  et al. }
   \titlerunning{Observation of EGRET Gamma-ray sources ...}

   \maketitle
%

\section{Introduction}
EGRET instrument on-board $Compton \hspace{1mm} Gamma\hspace{1mm}
Ray \hspace{1mm}Observatory\hspace{1mm}(CGRO)$ has detected both
diffuse and discrete gamma-ray emission. The diffuse emission is
both galactic (\cite{hunter}) and extra-galactic in nature
(\cite{sreekumar}). EGRET has detected about 271 high energy
($>$100 MeV) gamma-ray sources (\cite{hartman}). Besides AGNs
these sources include 170 sources that are not identified
conclusively with unique counterparts in other wavelengths. Two
third of these EGRET unidentified (EUI) sources lie close to the
galactic plane-potential counterparts (\cite{bhatacharia}) for
these include young pulsars, young radio quiet pulsars
(\cite{torres}, \cite{d'amico}, \cite{zhang}), Wolf Rayet (WR)
stars, Of stars, OB associations (\cite{Romero}), Super Nova
Remnants (SNRs) (\cite{combi}, \cite{case}, \cite{sturner}) and
other types  of sources.\\ Some other faint sources are in the
mid-latitude region suggested to be associated with the Gould Belt
(\cite{Gehrels}), which underwent an intense star formation period
about sixty million years ago (\cite{grainer}, \cite{harding}).
High latitude sources which are about 50, might be galactic
gamma-ray halo sources (\cite{dixon}) or unidentified sources are
thought to be extra-galactic. These extra-galactic EUI sources
contain Blazars and Active Galactic Nuclei (AGNs), galaxy clusters
(\cite{colafrancesco}), BL Lacerta objects (\cite{torres2003}) and
other types.\\ Whether the EGRET sources have emission in higher
energies, is an interesting question (\cite{lamb}). Gamma-ray with
energies about 100 TeV and more entering the earth atmosphere,
produce Extensive Air Showers (EASs) (\cite{gaisser}) which could
be observed by the detection of the secondary particles of the
showers on the ground level (\cite{PhDbahmanabadi}). Previous
attempts have been reported by other EAS arrays (\cite{tibet},
\cite{tibet2000}, \cite{Borione}, \cite{Alexandreas},
\cite{casamia} ).\\ This paper reports the results of a small
particle detector array located at the Sharif University of
Technology in Tehran. This small array is a prototype for a larger
EAS array to be built at an altitude of 2600 m ($\equiv$ 756
g/cm$^2$) at ALBORZ Observatory (AstrophysicaL oBservatory for
cOsmic Radiation on alborZ) (\cite{observatory}) near Tehran. The
prototype installed on the roof of physics department of Sharif
University of Technology in Tehran, 1200 m ($\equiv$ 890
g/cm$^2$), $35.72^{\circ} N$ and $51.33^{\circ} E$. In this work
we present the observational results of 10 EGRET third catalogue
sources, we describe the experimental setup in Section 2. the data
analysis in Section 3, the results in Section 4. Section 5 is
devoted to a discussion of the results.
\section{Experimental arrangements}

\begin{figure}
   \centering
   \includegraphics[height=16.2cm,width=17.4cm]{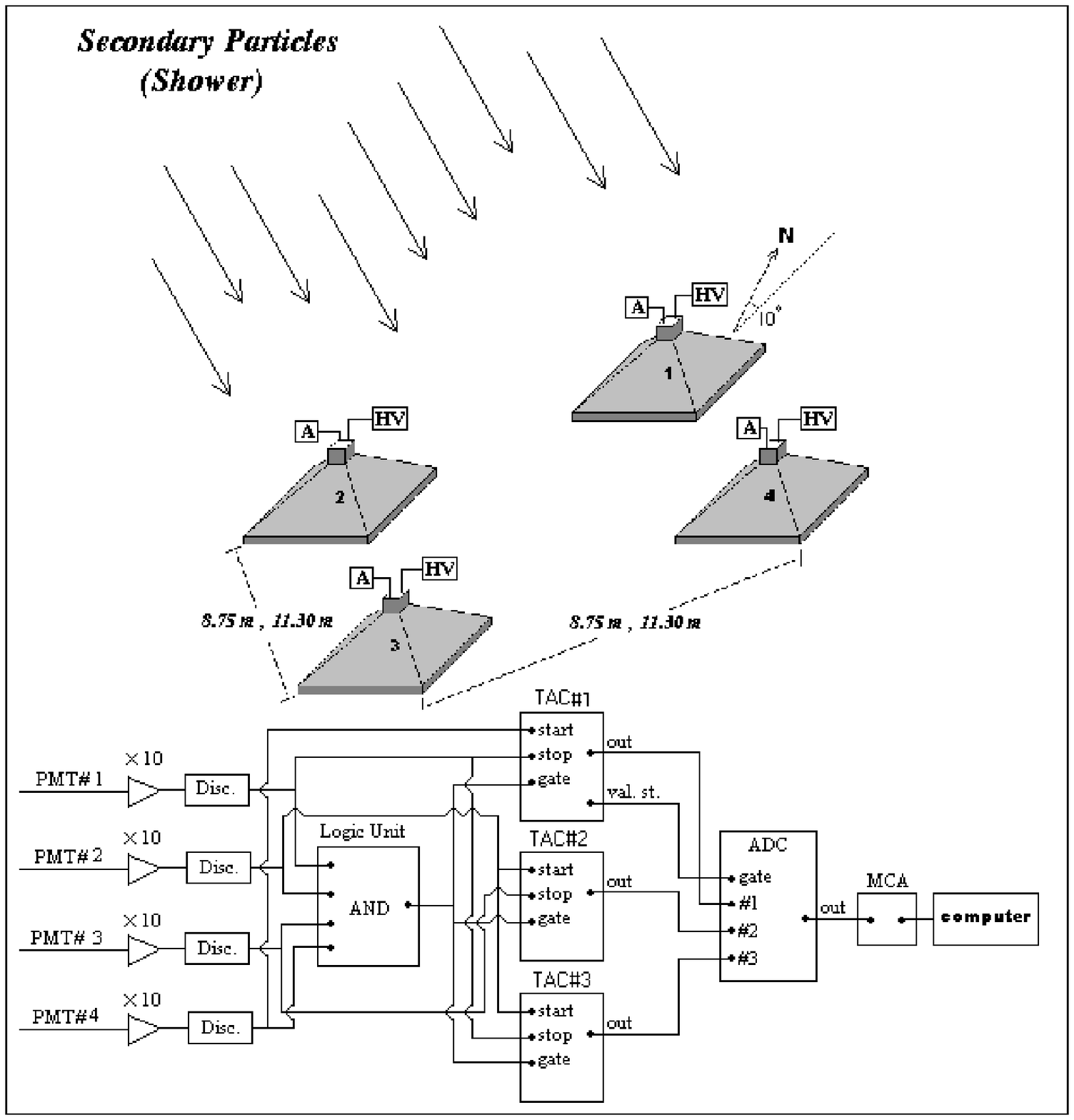}
      \caption{Experimental set up and electronic circuits.}
         \label{setup}
   \end{figure}

The array is constructed of 4 slab plastic scintillators
($100\times100\times2$ cm$^3$) as a square in Tehran (35$^{\circ}$
43$^{'}$ , 51$^{\circ}$ 20$^{'}$), Iran, with The elevation 1200 m
over sea level (890 g/cm$^2$) which is shown in Fig.~\ref{setup}.
All of the scintillators are on a flat level surface. Each
scintillator is housed in a pyramidal steel box with height of 15
cm. The interior surface of each box is coated with white paint,
(\cite{PhDbahmanabadi}) and a 5 cm diameter PMT(EMI 9813KB) is
placed at the vertex of the pyramidal box. Fig.~\ref{setup} shows
a schematic diagram of the array and its electronic circuit to log
each EAS event. After passing of at least one particle from a
detector the PMT creates a signal with a pulse height which is
related to the direction, number of the passed particles, and
location of the crossed particles in the scintillator. The output
signals from the PMTs are amplified in a one stage amplification
($\times$10) with an 8-fold fast amplifier (CAEN N412), and then
transfer to an 8-fold fast discriminator (CAEN N413A) which is
operated in a fixed level of 20mV one by one. The threshold of
each discriminator is set at the separation point between the
signal and background noise levels. Each discriminator has two
outputs, one of them is connected to a coincidence logic unit
(CAEN N455) as trigger condition. Trigger condition is satisfied
when at least one charged particle passes through each of the four
detectors within a time window of 150ns. The other discriminator
output is connected to a Time to Amplitude Converter (TAC)(EG\&G
ORTEC 566) which are set to a full scale of 200ns (maximum time
difference between each two scintillators which is acceptable).
The outputs of the No.4 scintillator was connected to start input
of TAC1 whereas the output of No.2 was connected to start inputs
of TAC2 and TAC3. The Output of the scintillator No.3 was
connected to the stop input of TAC2 and No.1 was connected to stop
inputs of both TAC1 and TAC3. Then the outputs of these three TACs
were fed into a multi parameter Multi Channel Analyzer (MCA)(KIAN
AFROUZ Inc.) via an Analogue to Digital Converter (ADC)(KIAN
AFROUZ Inc.)
unit.\\
When all of the scintillators have coincidence pulses, these TACs
are trigged by logic unit and 3 time lags between the output
signals of PMTs (4,1), (2,3) and (2,1) are read out by a computer
as parameters 1 to 3. So with this procedure an EAS event is logged.\\
Two different experimental configurations were used by the
experimental set up. All of the experimental set up were identical
in the first ($E1$) and the second ($E2$) experimental
configurations except for the size of the array. In $E1$ the size
was 8.75 m $\times$ 8.75 m and in $E2$ the size was $11.30$ m
$\times$ $11.30$ m.
\section{Data Analysis}

The logged time lags between the scintillators and Greenwich Mean
Time (GMT) of each EAS event were recorded as raw data. We
synchronized our computer to GMT
(\cite{http://www.timeanddate...}). Our electronic has record
capability of 18.2 times per second or equivalently each 0.055
seconds one record will be stored regardless of the existence or
non existence of EAS events. If an EAS event occurs, its three
time lags will be recorded and if it does not occur 'zero' will be
recorded. Therefore with the starting time of each experiment and
counting of these zero and non zero records we will obtain GMT
time of each EAS event. Our detected EAS events are a mixture of
cosmic-ray events and gamma-ray events. In $E1$ total number of
EAS events was 53,907 and duration of the experiment was 501,460
seconds. So the mean event rate of the first experiment was 0.1075
events per second. The distribution of the time between successive
events has a good agreement with an exponential function,
indicating that the event sampling is completely random
(\cite{submitted}). In $E2$ total number of events was 173,765 and
duration of the second experiment was 2,902,857
seconds. So its mean event rate was $0.05986$ events per second.\\
We refined the data for separation of acceptable events. Events
are acceptable if there be a good coincidence between the four
scintillator pulses. We omitted the events with zenith angles more
than 60$^{\circ}$. Therefore after the separation we obtained
smaller data sets of 46,334 and 120,331 for $E1$ and $E2$
respectively. Since we can not determine the energy of the showers
on an event by event basis, we estimate our lower energy threshold
by comparing our event rate to  a cosmic-ray integral spectrum
(\cite{Borione})
\begin{equation}
J(E) = 2.78\times10^{-5}E^{-2.22} + 9.66\times10^{-6}E^{-1.62} -
1.94\times10^{-12} \hspace{3mm} 40\leq E \leq 5000
\hspace{1mm}{\rm TeV}
\end{equation}
The obtained lower energy limits were 39 Tev in $E1$ and 54 TeV in
$E2$. The calculated mean energies were 94 and 132 TeV in $E1$ and
$E2$ respectively. If the well-known Hillas spectrum
(\cite{gaisser})
\begin{equation}
F(>E) \sim
2\times10^{-10}\frac{particle}{cm^2\hspace{1mm}s\hspace{1mm}sr}\times(\frac{E}{1000TeV})^{-\gamma}
\end{equation}
 is used the lower limits will be 40 TeV and 60
Tev. Since the distribution of cosmic-ray events within the array
in these energy ranges are homogeneous and isotropic, we used an
excess method (\cite{tibet}, \cite{tibet2000}) to find signature
of EGRET third catalogue gamma-ray sources. This method was used for both $E1$ and $E2$.\\
The complete analysis procedure is itemized as follows :
\begin{itemize}
 \item The calculation of local coordinates; zenith and azimuth angles of each EAS event $(z,\varphi)$
 were calculated using a least square method by
 logged time lags and coordinates of the scintillators.
 \item The local angle distributions of the EAS events were investigated to understand the
 general behaviours of these EAS events.
 \item The calculation of equatorial coordinates (RA,Dec) of each EAS event
 using its local coordinates, the GMT of the event and geographical
 latitude of the array. Then we calculated galactic coordinates (l,b) of each EAS event
 from its equatorial coordinates using epoch J2000.
 \item The estimation of angular errors in galactic coordinates of investigated EGRET sources
 by error factors of the array.
 \item The simulation of a homogeneous distribution of EAS events to
 investigate cosmic-ray EAS events.
  This simulation incorporated all known parameters of
 the experiment.
 \item The investigation of the statistical significance of random sources and
 the significance of sources from the third EGRET catalogue using the method
 of Li \& Ma (\cite{li_Ma})  and find the best location for EGRET
 sources in the TeV range.
 \end{itemize}
\subsection{Calculation of Local coordinates of each EAS event}

The local coordinates are zenith $(z)$ and azimuth $(\varphi)$.
We used the least square method (\cite{mitsui}) to calculate $z$
and $\varphi$. It is assumed that the shower front could be
approximated
by a plane. So we obtain,\\
\begin{equation}
\tan(z)=\sqrt{\frac{X^2+Y^2}{1-X^2-Y^2}}\hspace{5mm},\hspace{5mm}\tan(\varphi)=Y/X
\end{equation}
where,
\begin{eqnarray}
 X=\hspace{1mm}c\hspace{1mm}\vline
\begin{array}{cc}
  \sum x_{oj} t_{oj} & \sum x_{oj} y_{oj} \\
  \sum y_{oj} t_{oj} & \sum y^2_{oj}
\end{array}\vline \hspace{1mm}/ \hspace{1mm}\vline\begin{array}{cc}
  \sum x^2_{oj} & \sum x_{oj} y_{oj} \\
  \sum x_{oj} y_{oj} & \sum y^2_{oj}
\end{array}
\vline\hspace{2mm},   \\
Y=\hspace{1mm}c\hspace{1mm}\vline\begin{array}{cc}
  \sum y_{oj} t_{oj} & \sum x_{oj} y_{oj} \\
  \sum x_{oj} t_{oj} & \sum x^2_{oj}
\end{array}\vline\hspace{1mm} / \hspace{1mm}\vline\begin{array}{cc}
  \sum x^2_{oj} & \sum x_{oj} y_{oj} \\
  \sum x_{oj} y_{oj} & \sum y^2_{oj}
\end{array}\vline\hspace{2.5mm}.
\end{eqnarray}
\\

$\bf{D}_{oj}=\bf{D}_j-\bf{D}_o$=$x_{oj}\hat{\bf{i}}$+$y_{oj}\hat{\bf{j}}$
and $t_{oj}=t_j-t_o$ are the coordinate vector and the time
lag of j$_{th}$ scintillator with respect to the reference one and $c$ is the velocity of light.\\
A zenith angle cut off $60^{\circ}$ is implemented to enhance
significance.

\subsection{Angular distribution of the EAS events}

\begin{figure}
   \centering
   \includegraphics[height=14.9cm,width=10.1cm]{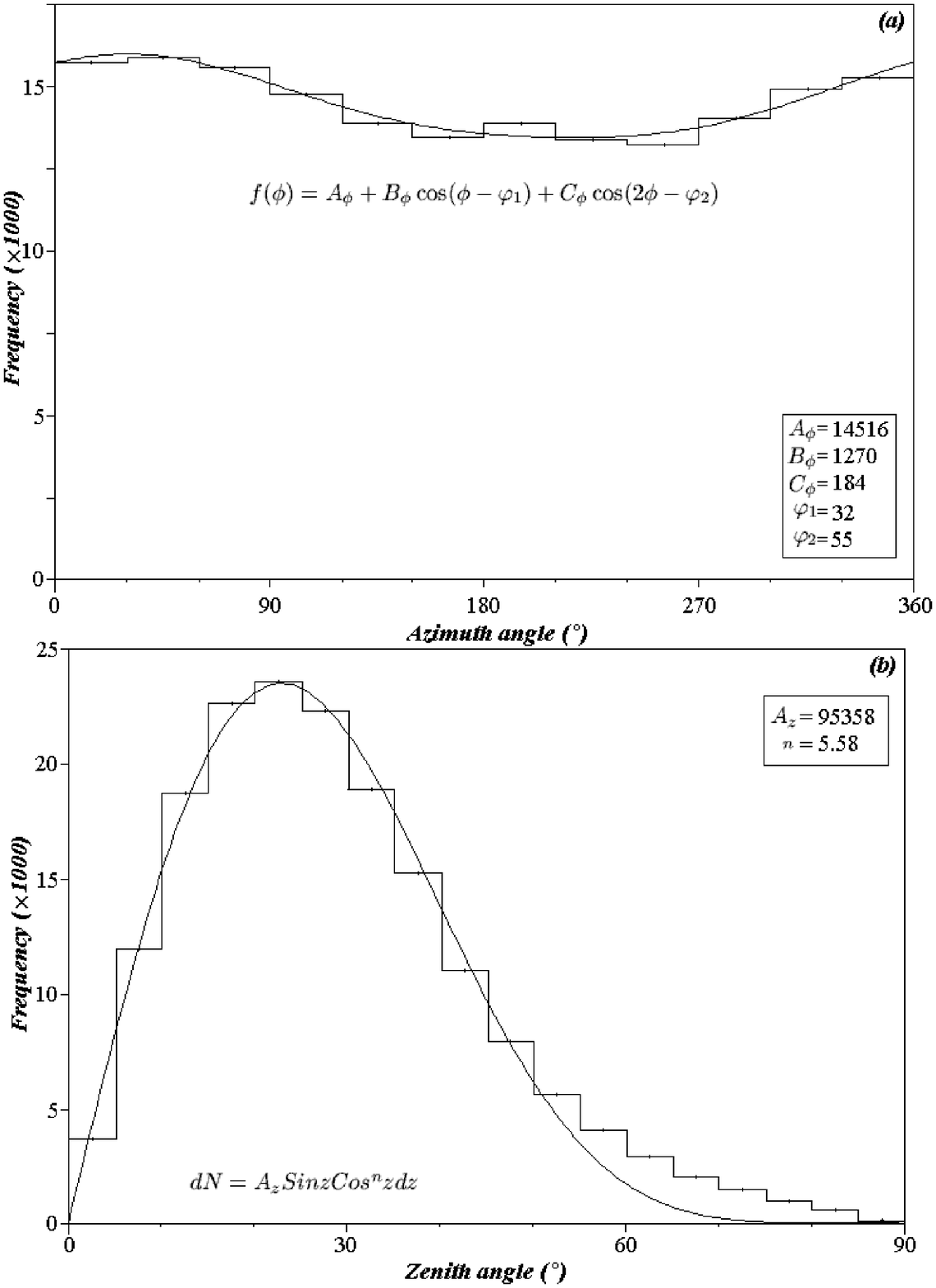}
      \caption{Local coordinates distributions of, (a) azimuth '$\phi$' and (b) zenith '$z$' angles of logged EAS events.}
         \label{ttfi}
   \end{figure}

Fig.~\ref{ttfi}(a) shows the azimuthal angle distribution of the
EAS events which is nearly isotropic. A slight North-South
anisotropy is observed which is attributed to the geomagnetic
field. We fitted this distribution with a harmonic function as
follow : (\cite{magnetic})\\
\begin{equation}
\\f(\phi)=A_{\phi}+B_{\phi}\cos(\phi-\varphi_1)+C_{\phi}\cos(2\phi-\varphi_2)
\end{equation}
where A$_{\phi}$, B$_{\phi}$ and C$_{\phi}$ are respectively
14516, 1270 and 184. $\varphi_1$ and $\varphi_2$ are phase
constants which are respectively 32$^{\circ}$ and 55$^{\circ}$.\\
Since thickness of the atmosphere increases quickly with
increasing zenith angle $z$ (\cite{gaisser}), the number of EAS
events is strongly
related to the $z$ value, as shown in Fig.~\ref{ttfi}(b).\\
These distributions were studied separately for the two
experimental configuration $E1$ and $E2$. The shower rate in $E2$
is less than $E1$ because of the larger size of the array in $E2$.
However the zenith angle distributions in $E1$ and $E2$ are very
similar. The differential zenith angle distributions of these data
sets are fitted to the function $dN=A_{z}SinzCos^nzdz$ with a very
good agreement for both $E1$ and $E2$ which from 0$^{\circ}$ to
50$^{\circ}$ $A_z=95358$ and $n=5.85$ and from 50$^{\circ}$ to
60$^{\circ}$ $A_z=90189$ and $n=5.00$. From another view the mean
value of zenith angle $(\bar{z})$ is $26.2746$ and $26.4625$ in
$E1$ and $E2$ respectively. Since the results of the two
experimental configurations, are in a good agreement with one
another and the excess is important for us. Therefore we added the
two data sets to obtain a larger data set with lower energy
threshold of $E1$ which is 39 TeV.

\subsection{Calculation of equatorial and galactic coordinates of each EAS event}

The equatorial coordinates (RA,Dec) are obtained from calculated
local coordinates ($z,\varphi$), GMT of each EAS event and
geographical latitude of the array. In this step the
transformation relations (\cite{coordinate}, \cite{osouloamal}),
and the local sidereal time of the starting point of the
experiment (\cite{tycho}) were used.\\
Then galactic coordinates (l,b) of each EAS event are obtained
from the calculated equatorial coordinates, based on the galactic
coordinate standard of year 2000 (\cite{coordinate}).
Fig.~\ref{dat} shows the distribution of our data in galactic
coordinates.
\subsection{Error estimation of investigated sources in galactic coordinates}

For the coordinates calculations of each EAS event in galactic
coordinates we have to know estimated errors in these coordinates.
These errors are due to experimental error factors, which contain
uncertainties in time and coordinates of each logged EAS event.
The defined distance between two scintillators was centre to
centre and the size of the scintillators were
($100\times100\times2$ cm$^3$). Meanwhile the accuracy of
coordinates of each scintillator is measured within a few
centimeters. So error in measurement of coordinates of secondary
particles of each EAS event is $\Delta d\cong1$ m.\\
The errors in time measurement of each EAS event are due to the
front thickness of the secondary particles, electronics errors and
error in GMT logging. The error due to the first two factors was
$\Delta t\cong2$ns (\cite{magnetic}). The error in GMT logged time
of each EAS event was $\Delta T=0.07$s which is due to recording
rate and the synchronizing of the computer. These errors make
uncertainties in galactic coordinates of the
investigated sources by the array.\\
The following errors were calculated :\\
\begin{itemize}
 \item The errors in local, equatorial and galactic coordinates of each EAS event.
 \item The observational angular error of each source.
 \item The mean and standard deviation of these error angles. We
 calculated these steps
  for more than 1000 random sources which are in the
Field Of View (FOV) of the array. This calculation was carried out
for $E1$ and $E2$ separately and was weighted by their refined EAS
events.
\end{itemize}
From geometry of Fig.~\ref{setup} we can drive :
\begin{eqnarray}
\sin z \sin\varphi = \frac{c}{d}(t_2 - t_1)\\
\sin z \cos\varphi =\frac{c}{d}(t_3 - t_2)
\end{eqnarray}
In these relations $t_i$s are logged times of an EAS event in
i$_{th}$ component of the array and $d$ is side of the square array.\\
The errors in zenith and azimuth angles were obtained by
differentiating from eqs. (5) and (6) :
\begin{equation}
\Delta z^2 = A_{\Delta}^2 / \cos^2 z + B_{\Delta}^2 \tan^2 z
\end{equation}
\begin{equation}
\Delta \varphi^2 = A_{\Delta}^2 / \sin^2 z
\end{equation}
where $A_{\Delta}=2c\Delta t/d$ and $B_{\Delta}=\Delta d/d$. The
errors in equatorial and galactic coordinates were calculated from
differentials of Dec(z,$\varphi$), RA(z,$\varphi,T$), b(RA,Dec)
and l(RA,Dec).\\If $y$ is a generic function of parameters $u$,
$v$ and $T$, then :
\begin{equation}
y=y(u,v,T)\\
\end{equation}
\begin{equation}
|\Delta y| = \sqrt{(\frac{dy}{du})^2 \Delta u^2+
(\frac{dy}{dv})^2 \Delta v^2 + (\frac{dy}{dT})^2 \Delta T^2}
\end{equation}
Where $dy/dT=2\alpha$ ($\alpha =1.00273790935$) (\cite{tycho}) for
the calculation of RA(z,$\varphi,T$) and $dy/dT=0$ for
calculation of Dec(z,$\varphi$),
b(RA,Dec) and l(RA,Dec).\\
The error on the observed solid angle of each source is $\Delta
\Omega = \cos$b$\Delta$b$\Delta$l and the equivalent error on the
angular radius is $r_e\cong \sqrt{\Delta\Omega/\pi}$
($r_e \ll 1 $)\\
The upper analysis obtains angular resolution of each EAS event
individually. In case that there are many EAS events with
different local coordinates which have contributions in the
signature of each investigated source. Therefore at first angular
errors of all of the accumulated EAS events in the galactic
coordinates of the source were calculated, then the mean value of
these angular errors was chosen as angular error of the source for
the first step. Since all of the accumulated EAS events in the
angular error region have contributions in the source signature,
so the previous calculations were repeated for all of the
accumulated EAS events in a circular region with the center of the
source and radius of $r_e$. Finally the mean value of these EAS
angular errors is calculated as angular error of each source in
galactic coordinates. Since the side distances of the array is
different in $E1$ and $E2$, angular errors in these two
experiments are different. So for calculation of the final result
for each source these angular errors calculated separately for
$E1$ and $E2$ and was weighted with the number of refined EAS
events in the related experiment. The final angular errors of
investigated sources ($r_e$) are shown in Table 1. Since these
angular error radii have a little fluctuations respect to their
mean, therefore we sampled over l and b with a step of 5 degrees
from the FOV of the array and calculated these radii to find the
mean and standard deviation. Therefore the mean value and the
standard deviation of the angular error of the experiment was
obtained from angular error of more than 1000 random points. With
These steps we obtained $\bar{r_e}=4.35^{\circ}\pm0.82^{\circ}$ as
the mean angular error of the experiment.
\subsection{Drawing exposure map and simulation of the experiment}
\begin{figure}
   \centering
   \includegraphics[height=17cm,width=8.5cm,angle=-90]{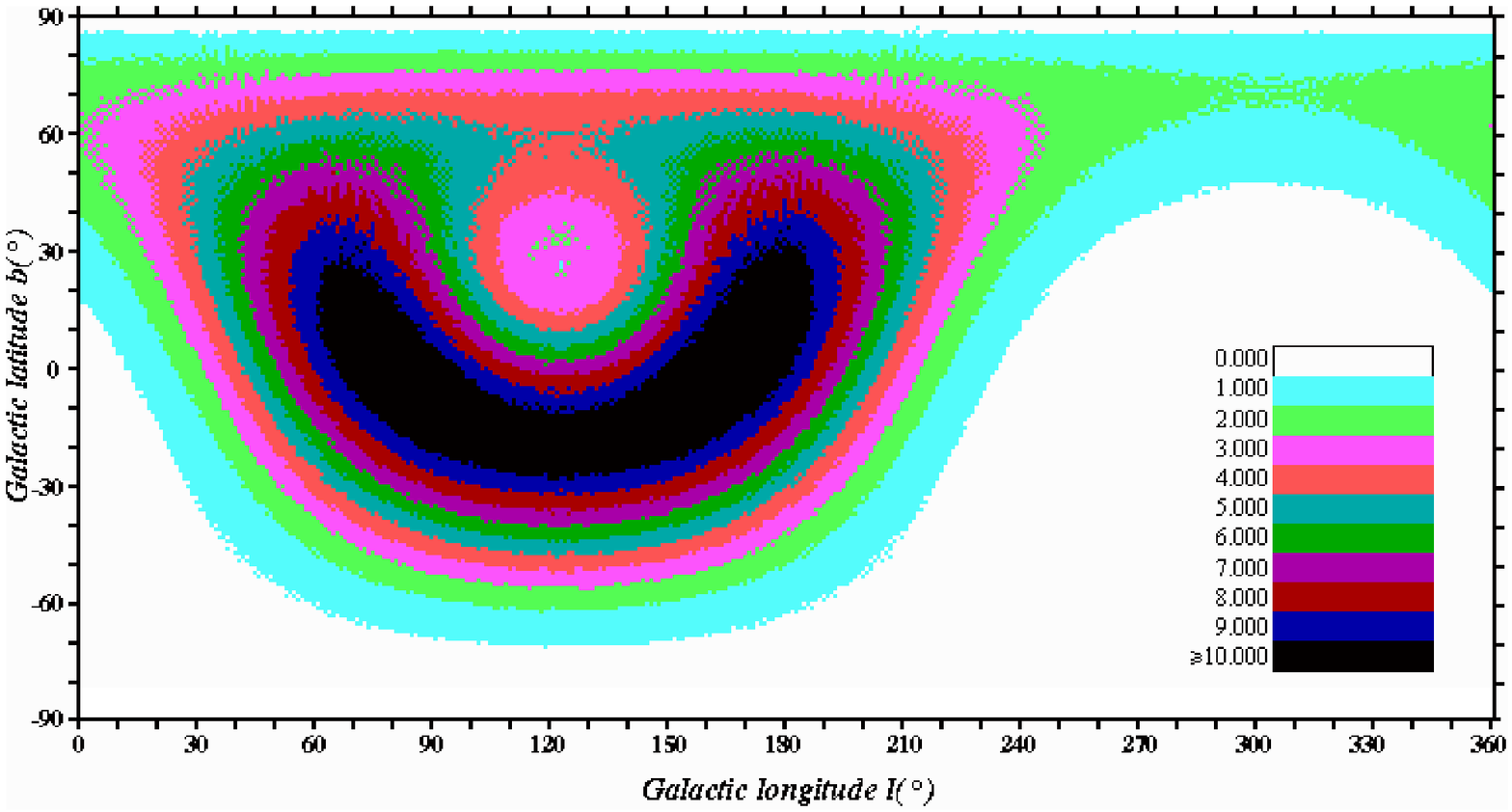}
      \caption{Exposure map of simulated events in 1$^{\circ}\times$1$^{\circ}$ bins base on the general parameters of
      total distribution of EAS events in galactic coordinates.}
         \label{sim}
   \end{figure}

Because of the various exposures of the sky in time, there is a
non-uniform distribution of EAS events in galactic coordinates.
The variation in time exposures due to the altitude difference of
different sources, and the observation of separate individual
galactic regions during the sub-intervals within the long duration
of the experiment were simulated. We have 166665 EAS events in our
experiments, so we used monte carlo method for this simulation and
we simulated 166665 random events. These random numbers were
chosen with considerations of Fig.~\ref{ttfi}. From this figure is
seen that $\phi$ distribution is not isotropic and the thickness
effect of the atmosphere is very important, these effects were
considered in choosing of these random numbers. So in the
procedure :
\begin{itemize}
\item Zenith angle $(z)$ was taken from 1$^{\circ}$ to
60$^{\circ}$.
\item Azimuth angle $(\phi)$ was chosen from 1$^{\circ}$ to
 360$^{\circ}$.
 \item Related random numbers of time were chosen with
consideration of EAS event rate of the experiment. Meanwhile we
considered duration of the experiment which was taken from start
and stop times of each sub-experiment.
\end{itemize}
With this procedure we obtained 2500 simulated map and obtained
the map with mean number of simulated events per
($1^\circ\times1^\circ $) pixel with the accuracy of 0.001.
Fig.~\ref{sim} shows the exposure map of the experiment. The event
map in Fig.~\ref{dat} reflects the uneven exposure of the
experiment.
\subsection{Investigation of EGRET gamma-ray sources and measurement of their statistical significance}

\begin{figure}
   \centering
   \includegraphics[height=18cm,width=10.1cm]{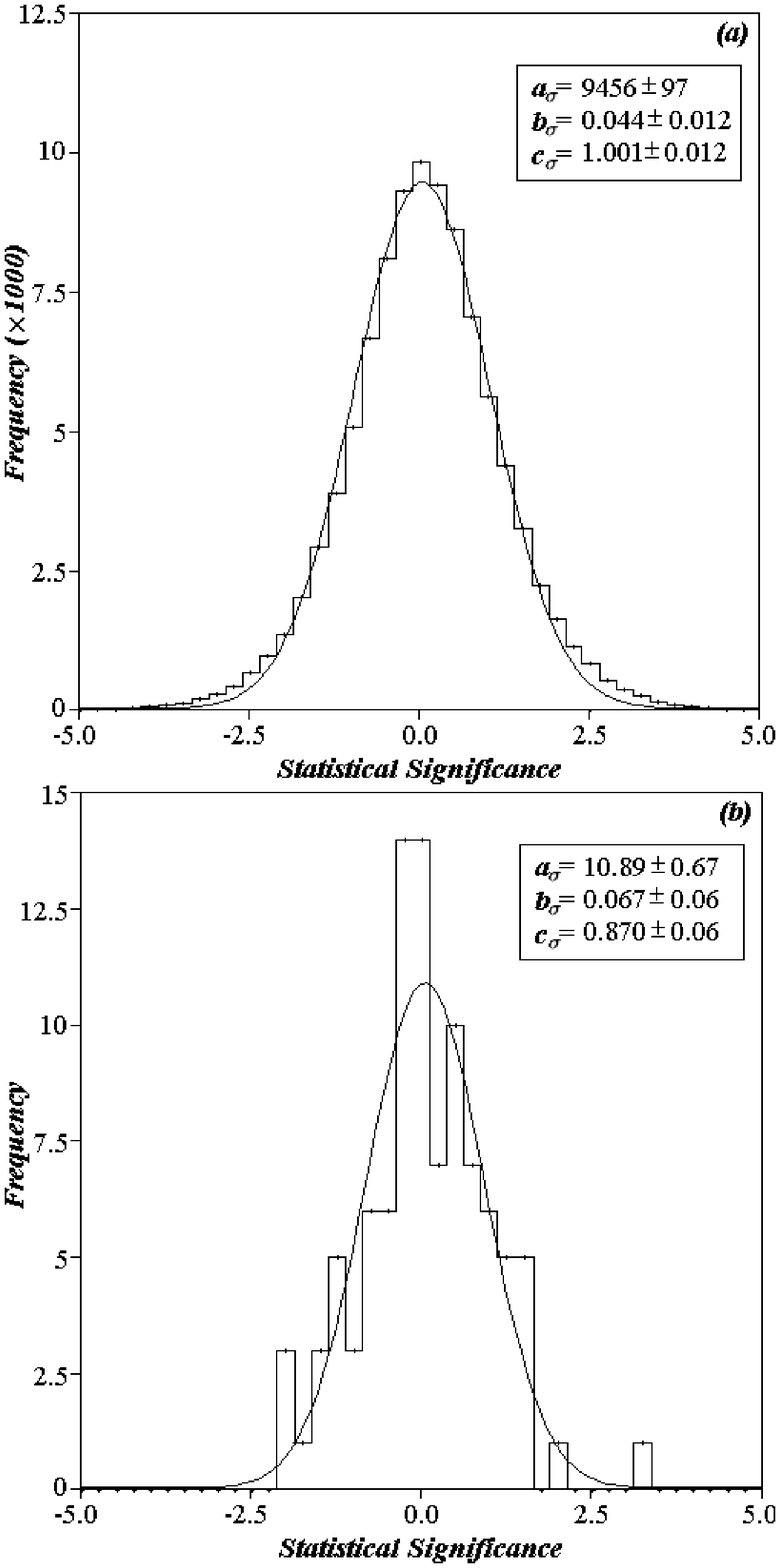}
      \caption{Distribution of frequency of (a) 98,000 virtual random sources and (b) 98 EGRET
      sources in the FOV of our array in galactic coordinates versus their statistical
      significance. a$_{\sigma}$, b$_{\sigma}$ and c$_{\sigma}$ are described in eq. (14)}
         \label{dst}
   \end{figure}

The energy range of the logged EAS events by the array is in the
range of 40 to 10,000 TeV. In this energy range distribution of
cosmic-rays is completely isotropic and homogeneous in the galaxy.
After correcting for the exposure effects, we looked for excess
emission that could be from gamma-ray sources. We used third EGRET
catalogue (\cite{hartman}) as a reference. But some of EGRET
sources have not acceptable events in the FOV of our array. So we
counted number of events, number of pixels and then calculated
count per pixel related to each of these sources. We note that the
mean count per pixel in data map is 4.798. Of 151 EGRET sources
only 123 of them have count per pixel of more than square root of
4.798 with 98 more than 1.5 times the square root of the mean. So
we started our investigations on these 98 sources. A method of
excess similar to the analysis adopted by the Tibet EAS array, has
been adopted (\cite{tibet} \& \cite{tibet2000}). In the first step
we divided the data map (Fig.~\ref{dat}) to the exposure map
(Fig.~\ref{sim}) pixel by pixel. in the obtained map,
approximately most of non zero pixels are around 1 except probable
source pixels and pixels with more fluctuations in the data map,
which probably due to the smallness of the data set. For
eliminating the fluctuated pixels we multiplied the new map to
4.798 as raw exposure corrected map. In this step we added counts
of all pixels of the raw corrected map. The number must be very
near to 166,665 so with this restriction we obtained a lower limit
0.0750 for eliminating pixels with less count in the exposure map,
and the final
exposure corrected map was obtained which is shown in Fig.~\ref{cor}.\\
The obtained map was fairly uniform in the FOV of our array in the
galactic coordinates. Next we investigated the remaining faint
inhomogeneity in the corrected map that could be conditionally
attributed to the existence of gamma-ray sources. To estimate the
significance of an individual source we obtained all corrected EAS
events, $N_{on}$, within a radius $\sqrt{2}r_e$from the source
position. The number of pixels, $n_s$, within this region was also
found. The total number of background counts, $N_{off}$, was found
from the pixels that fall within an outer radius of 2$r_e$ and an
inner radius $\sqrt{2}r_e$ from the source position. The number of
background pixels, $n_b$, was also calculated. The statistical
significance of the source was obtained using the Li \& Ma
relation (\cite{li_Ma}).
\begin{equation}
S =\frac{N_{on} - \alpha N_{off}}{\sqrt{N _{on}+\alpha^2
N_{off}}}\hspace{5mm} , \hspace{5mm}\alpha = \frac{n_s}{n_b}
\end{equation}
Distribution of statistical significance of these 98 sources is
fitted on a gaussian function as follow :
\begin{equation}
f(\sigma)=
a_{\sigma}exp(-\frac{(\sigma-b_{\sigma})^2}{2c_{\sigma}^2})
\end{equation}
which is shown in Fig.~\ref{dst}. Then for the investigation of
the statistical significance distribution, we chose 98,000 virtual
random sources with similar conditions of the 98 EGRET sources.
Fig.~\ref{dst} shows a normal distribution with mean 0.044 and
standard deviation 1.001. Procedure of the significance
calculation of these virtual sources is as like as the 98 EGRET
sources except a little difference. Basically the virtual sources
have no signals so we used another relation for the calculation of
their statistical significance (\cite{li_Ma}).
\begin{equation}
S =\frac{N_{on} - \alpha N_{off}}{\sqrt{\alpha(N_{on}+
N_{off})}}\hspace{5mm} , \hspace{5mm}\alpha = \frac{n_s}{n_b}
\end{equation}
\subsection{Investigation of a probable displacement of the
source signatures}

Our exposure corrected map has not any bright source signatures,
so we used third EGRET catalogue as a reference for searching some
sources in our energy range. But EGRET energy range is from 100
MeV to 30 GeV and our energy range is from 40 TeV to 10,000 TeV.
So to search for some sources in our data we ought to hope that
these sources have had a spread spectrum at least from EGRET
energy range to our energy range, like blazars, BL Lac objects,
Flat-spectrum radio quazars or etc. Since usually these sources in
different ranges of energies have not the same places exactly and
they have a little displacement, we searched around these sources
with one degree displacement. This displaced l and b are shown in
Table 1 for each source. It means that around each source with
statistical significance more than 1 we tried 8 ($1^\circ \times
1^\circ$)pixels around it and chose the location with highest
statistical significance.
\section{Results}

\subsection{Explanation of the Field of View (FOV) in galactic coordinates}

The rotation axis of the Earth passes near the star Polaris,
angular difference between Polaris and the rotation axis is
approximately 5 times smaller than our mean accuracy ($\bar{r_e}$)
in galactic coordinates. So in this analysis Polaris is considered
as being on the rotation axis of the Earth. The longitude and
latitude of Polaris in galactic coordinates are
$123^{\circ}$,$17^{'}$ and $27^{\circ}$,$28^{'}$ respectively. The
geographical latitude of Tehran is $35^{\circ}$ $N$, so the angle
between the zenith of the array and Polaris in Tehran is
$55^{\circ}$, we selected events with zenith angles less than
$60^{\circ}$ for the analysis which is deduced from
Fig.~\ref{ttfi}(b) and therefore, Polaris and regions around it
are observable only with High zenith EAS events. From
Fig.~\ref{ttfi}(b) it can be seen that the best observable region
is from $10^{\circ}$ to $40^{\circ}$ of zenith angles. In
Fig.~\ref{dat} is shown that Galactic longitudes smaller than
$l\approx123^{\circ}-(2 \times 60^{\circ} -
(60-55)^{\circ})\approx 8^{\circ}$ and larger than
$l\approx123^{\circ}+(2 \times 60^{\circ} -
5^{\circ})\approx238^{\circ}$ are less observable. In other words,
given the location of the array these are two different observable
regions in galactic coordinates. Galactic latitudes smaller than
$b\approx27^{\circ}-(60^{\circ} - 5^{\circ})\approx-38^{\circ}$
and larger than
$b\approx27^{\circ}+(60^{\circ}-5^{\circ})\approx82^{\circ}$ are
less observable regions too.\\

\subsection{Comparison of observed sources of $E1$ and $E2$}

 With the procedure which is mentioned in subsection 3.7. we
searched for sources with statistical significance more than 1.5,
and we found thirteen sources which five of them are more than 2.
For avoiding from probable fluctuations we did another try. we
searched these displaced sources in $E1$ and $E2$ separately. But
in this stage because of the smallness of these data sets,
specially in $E1$ we separated significance more than 1. Therefore
10 sources remained which have statistical significance more than
1 in $E1$ and $E2$ and more than 1.5 in the sum, which are shown
in Table 1. Fortunately five of these sources are AGNs, one is
probably AGN and four of them are unidentified sources. This is in
case that from 271 source of 3rd EGRET catalogue sources only 66
are AGNs.

\subsection{Distribution of observed shower events around most significant EGRET sources}

\begin{figure}
   \centering
   \includegraphics[height=14.9cm,width=10.1cm]{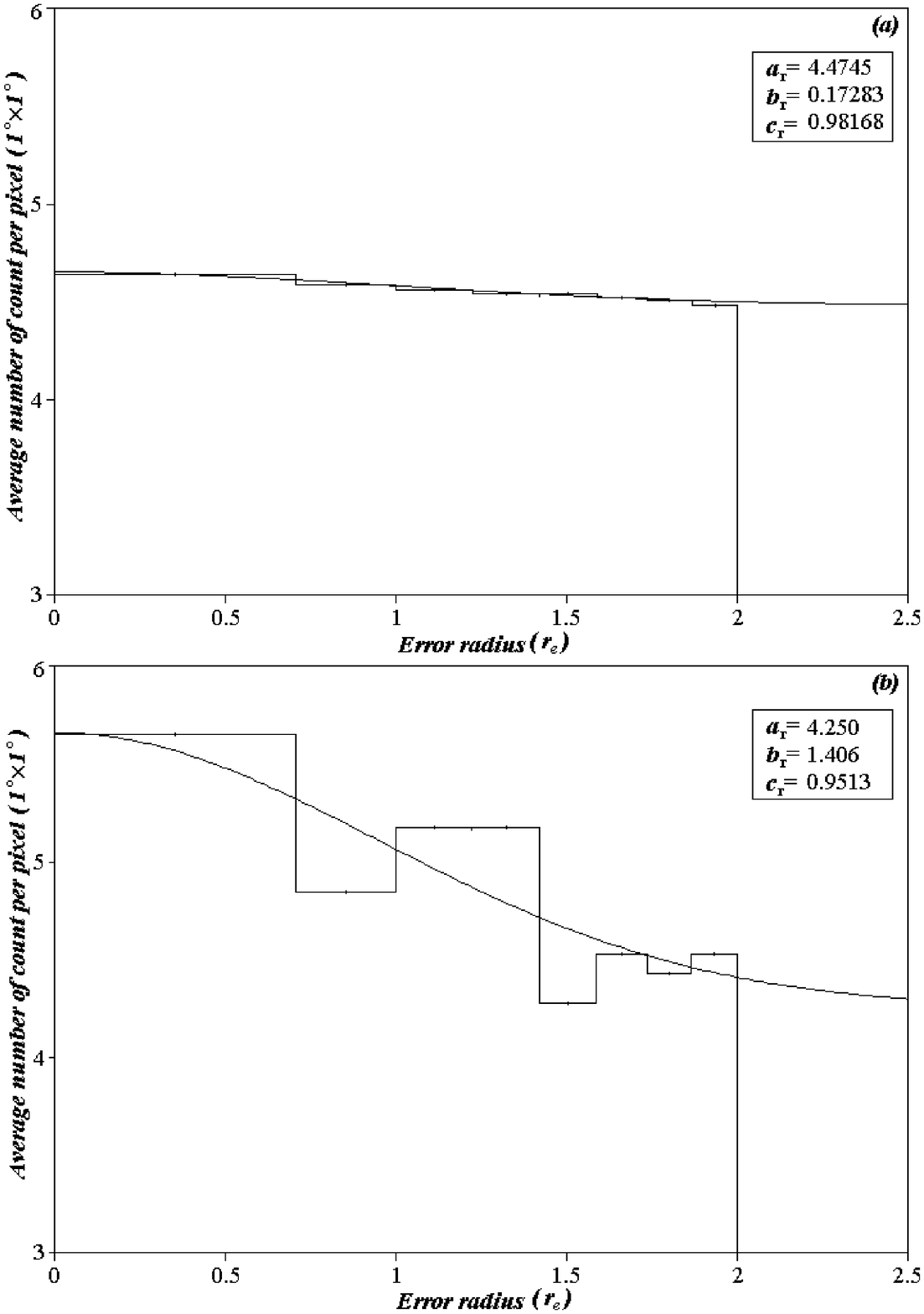}
      \caption{Distribution of mean count per pixel of (a) 98,000 virtual random sources
      and (b) 10 EGRET sources of Table 1 versus error radial distance from the centre of the related sources.}
         \label{radial_dst}
   \end{figure}

It seems that radial distribution of number of counts per pixel
for each source naturally must be near to a gaussian distribution
as a source signature, over a flat back ground. We separated eight
regions with approximately the same number of pixels for each
source. The first region is a circle with radius $\sqrt{1/2}r_e$.
The second region is a ring with inner radius $\sqrt{1/2}r_e$ and
outer radius $\sqrt{2/2}r_e$ and with this order we separated
eight regions. Distribution of mean counts per pixel around 98,000
virtual random sources and 10 most significant EGRET sources is
shown in Fig.~\ref{radial_dst}. These distributions fitted on a
gaussian function over a flat distribution as follow:
\begin{equation}
 f(r_e)=a_r+b_rexp(-r_e^2/2c_r^2).
\end{equation}
\section{Discussion and concluding remarks}
\begin{figure}
   \centering
   \includegraphics[height=17cm,width=8.5cm,angle=-90]{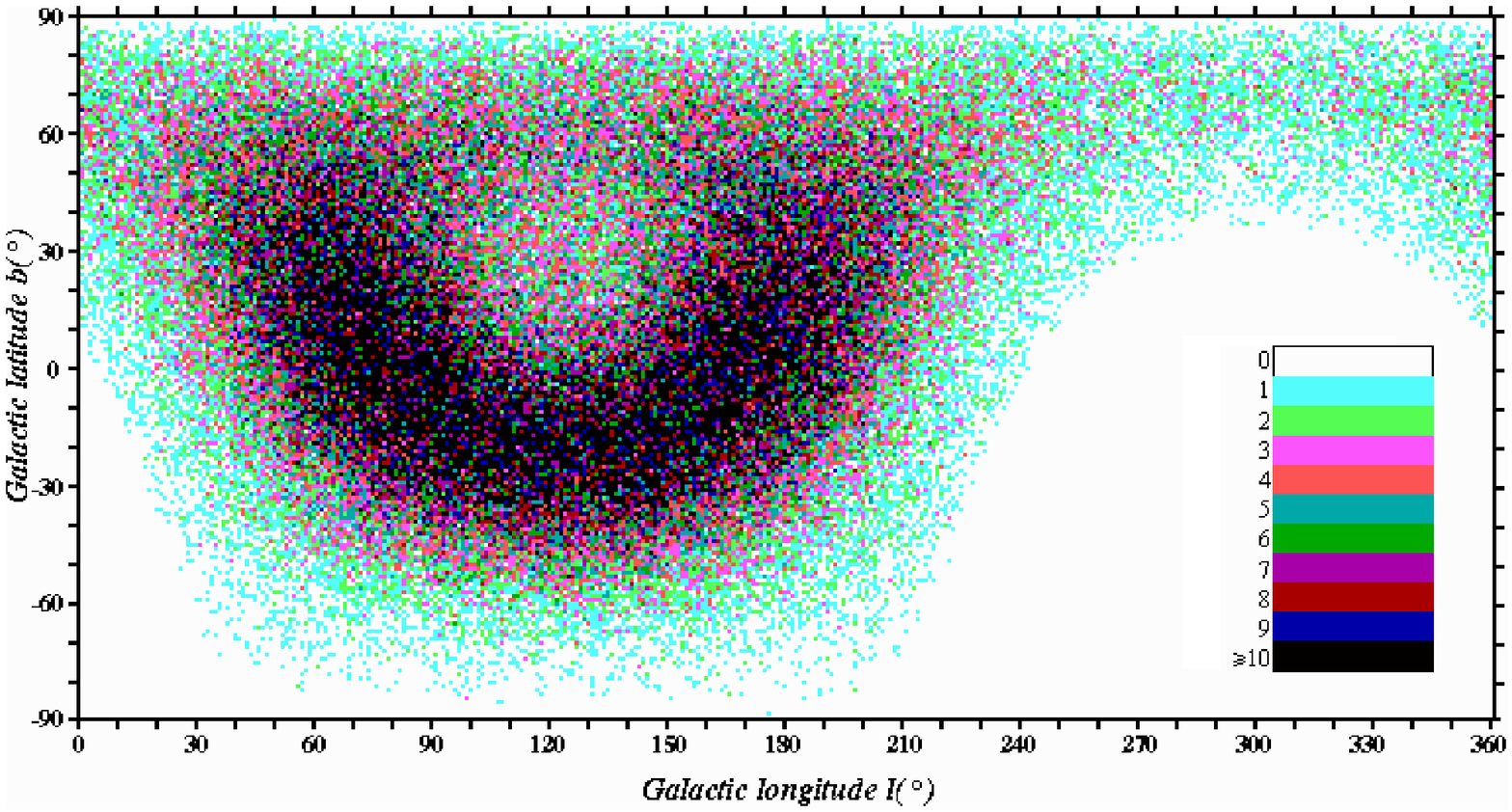}
      \caption{EAS events map in 1$^{\circ}\times$1$^{\circ}$ bin galactic coordinates.}
         \label{dat}
   \end{figure}

\begin{figure}
   \centering
   \includegraphics[height=17cm,width=8.5cm,angle=-90]{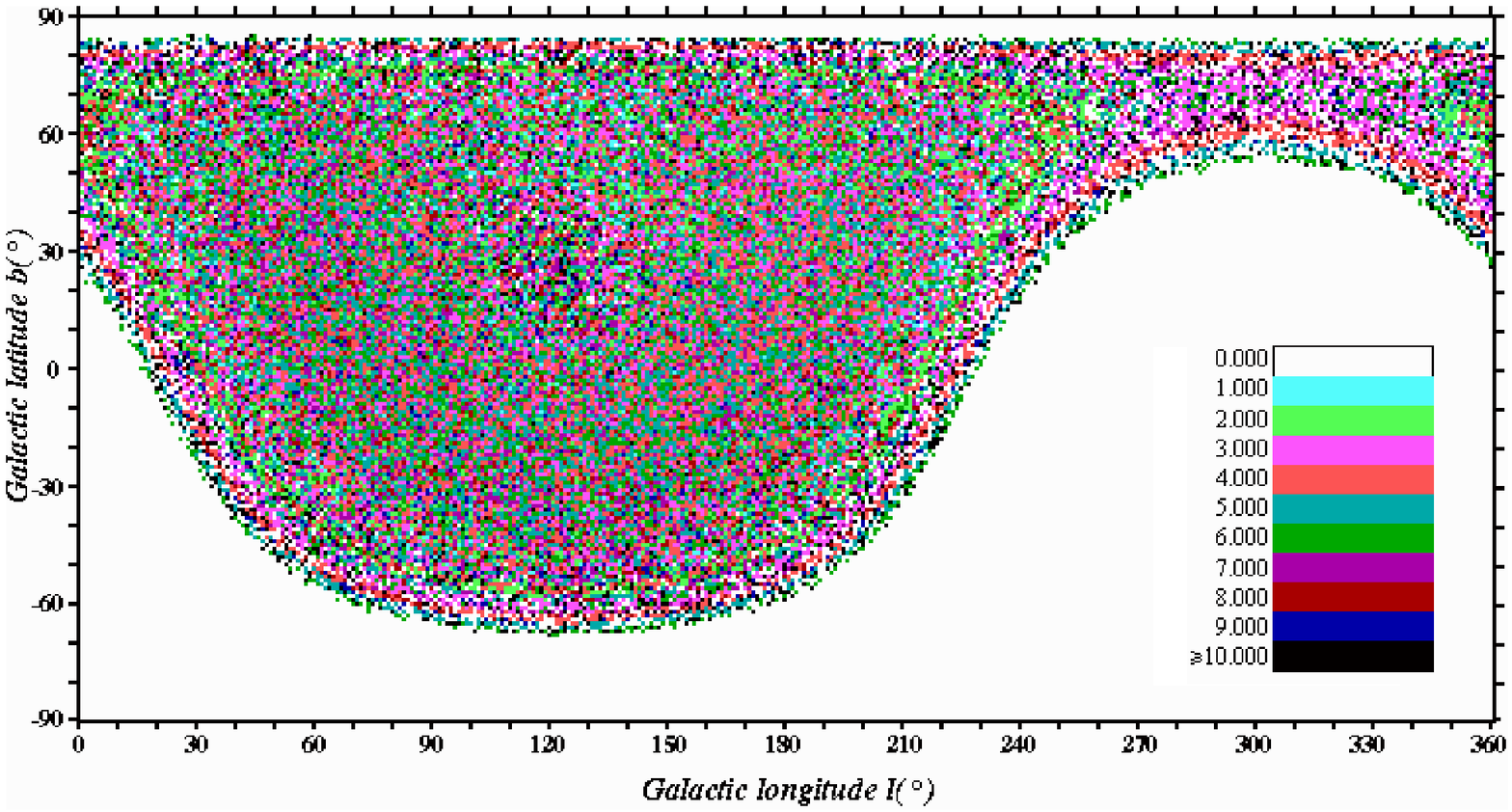}
      \caption{Corrected exposure map which is extracted from division of the data map (Fig.~\ref{dat}) to the
      exposure map (Fig.~\ref{sim}) pixel by pixel.}
         \label{cor}
   \end{figure}

\begin{figure}
   \centering
   \includegraphics[height=17cm,width=8.5cm,angle=-90]{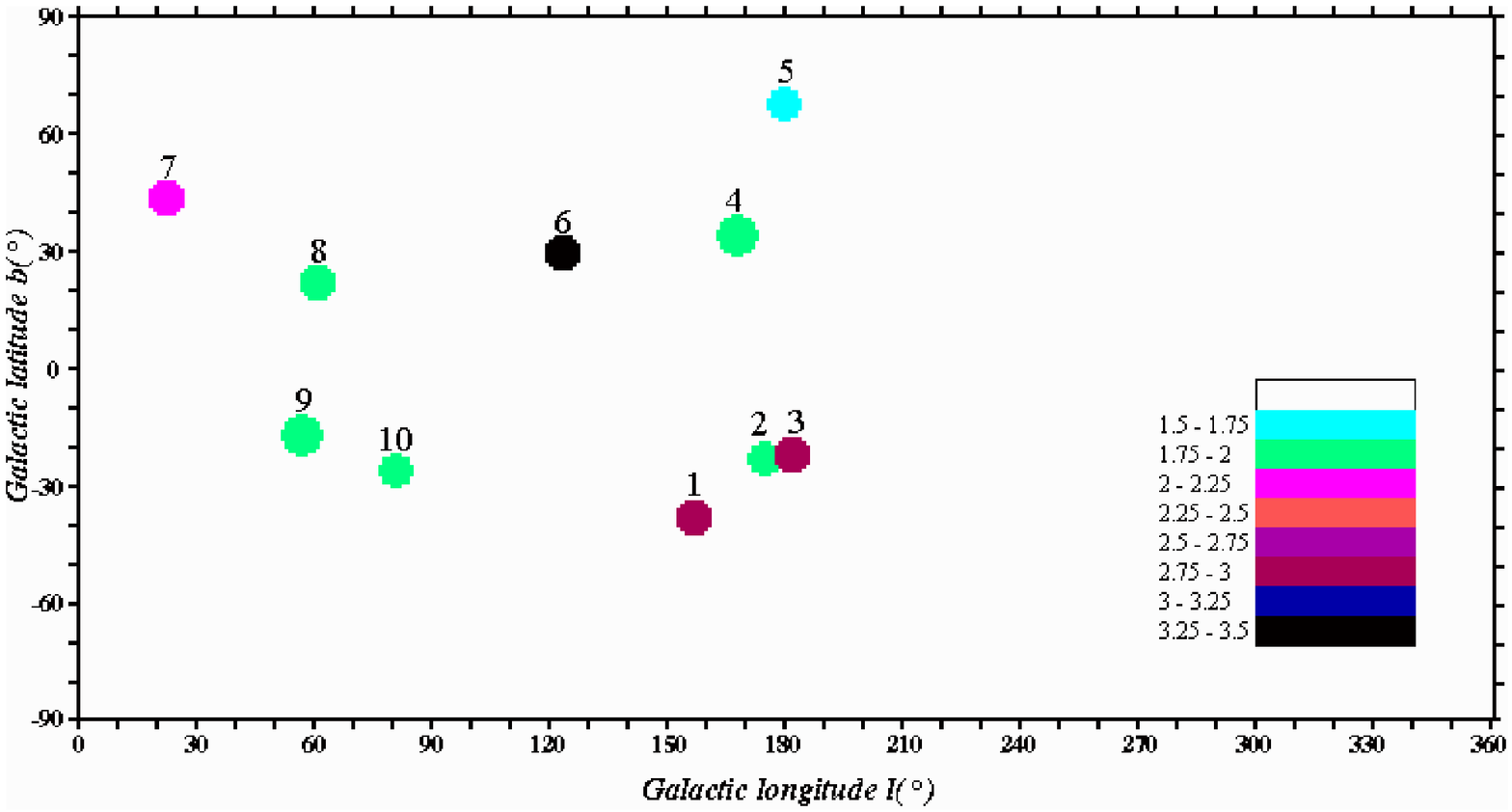}
      \caption{Map of EGRET sources with statistical significance more than 1.5 in galactic
      coordinates. The numbered sources were described in Table 1.}
         \label{sources}
   \end{figure}

In Table 1 is seen that most significant excesses observed, are in
the region $15^{\circ} < \bar{z} < 35^{\circ}$. This result is
reasonable because these angles are in favorable locations in the
sky and have considerably more data from this region. In addition
our data have a few counts in some parts of Fig.~\ref{dat} and we
were mandated to eliminate some source candidates from our list.
To increase the statistical significance of our results and
investigation of more sources, we have to accumulate
more data to have a map with less fluctuations.\\
There has been a considerable effort worldwide to detect gamma-ray
sources via the EAS technique. From a variety of arguments we
suspect that some, if not many, of the EGRET sources would be
detectable at very high energies. In this work, we are limited to
a discussion of a few sources with relatively small statistical
significance. Our statistical significance are not in a detection
limit with confident, we studied this procedure to guess some
candidates in unidentified EGRET sources more than TeV range. For
these sources listed in Table 1, we suspect that nine of them may
be extra-galactic ($|b|> 20^{\circ}$) (\cite{Gehrels}) and only
one is in galactic region ($|b|\leq 20^{\circ}$) and this one is
an AGN in the third EGRET catalogue list too. Four of our ten
sources were investigated before with CASA-MIA (\cite{catanese})
and two of them were GEV EGRET sources (\cite{lamb}). Therefore we
might expect that as many as four of these unidentified sources
could indeed be emitters at high energy and might be AGNs.\\ Some
of our observed sources overlap one another Fig.~\ref{sources}, so
a complete and accurate analysis procedure would incorporate the
maximum likelihood method (\cite{mattox}). We must also emphasize
that our experiment can not distinguish between gamma-ray and
cosmic-ray initiated air showers, and so we used the excess method
to carry out a search for very high energy gamma-ray emission.
After the analysis we understood that the record times per second
of our computer is very important and we have to increase this
record rate to decrease angular error radius of observable
sources. In our future site at 2600 m a.s.l. (\cite{observatory}),
we are constructing underground tunnels which will provide us with
ample space to deploy muon detectors. The detection of muons in
air showers should be provide a powerful away to discriminate
between cosmic-ray and gamma-ray air showers.

\begin{acknowledgements}
This research was supported by a grant from the national research
console of Iran for basic sciences.\\ The authors wish to thank
Dr. Dipen Bhattacharya at University of California, Riverside and
Prof. Rene A. Ong at University of California, Los Angeles for
their many constructive comments.\\ The authors wish to thank from
the anonymous referee for his/her many constructive comments too.
\end{acknowledgements}

\newpage
\begin{table}
  \begin{tabular}{ccccccccccccccc}
\hline\hline
    &Name&l&b&ID&l$_d$&b$_d$&$\sigma_{E1}$&$\sigma_{E2}$&$\sigma_{tot}$&$r_{e}(^\circ)$&$\bar{z}$&Flux&t1&t2\\
    \hline
    1 &3EG J0237+1635 & 156.46 & -39.28 &A& 157 & -39 & 1.29 & 2.80 & 2.90 & 4.70 & 24.87 & 630 &$\surd$&$\surd$\\
    2 &3EG J0407+1710 & 175.63 & -25.06 & & 175 & -24 & 1.65 & 1.28 & 1.95 & 4.78 & 24.05 & 782 &       &\\
    3 &3EG J0426+1333 & 181.98 & -23.82 & & 182 & -23 & 1.11 & 2.60 & 2.79 & 4.89 & 26.97 & 702 &       &\\
    4 &3EG J0808+5114 & 167.51 & 32.66  &a& 168 & 33  & 1.40 & 1.37 & 1.91 & 5.10 & 23.29 & 1241&       &\\
    5 &3EG J1104+3809 & 179.97 & 65.04  &A& 180 & 66  & 1.17 & 1.30 & 1.73 & 4.43 & 15.97 & 485 &$\surd$&$\surd$\\
    6 &3EG J1308+8744 & 122.74 & 29.38  & & 124 & 28  & 2.43 & 1.88 & 3.43 & 4.76 & 44.14 & 412 &       &\\
    7 &3EG J1608+1055 & 23.51  & 41.05  &A& 23  & 42  & 1.53 & 1.55 & 2.11 & 4.62 & 29.27 & 431 &$\surd$&\\
    8 &3EG J1824+3441 & 62.49  & 20.14  & & 61  &21   & 1.05 & 1.60 & 1.91 & 4.74 & 16.19 & 1370&       &\\
    9 &3EG J2036+1132 & 56.12  & -17.18 &A& 57  & -18 & 1.09 & 1.62 & 1.95 & 5.16 & 28.05 & 851 &       &\\
    10&3EG J2209+2401 & 81.83  & -25.65 &A& 81  & -27 & 1.45 & 1.60 & 1.86 & 4.25 & 21.40 & 844 &$\surd$&\\
  \end{tabular}
  \caption{Observed EGRET 3rd catalogue sources by our array. l$_d$ and b$_d$ are displaced galactic coordinates,
  $\sigma_{E1}$, $\sigma_{E2}$ and $\sigma_{tot}$ are  the statistical significance
related to the first experiment, the second experiment and sum of
both, $r_{e}$ is error angular radius, $\bar{z}$ is mean amount of
zenith angles of EASs related to each source and
  'Flux' is number of EAS events related to each source. t1 is AGNs which is investigated by CASA-MIA
  before (\cite{catanese}), t2 is sources with energy more than 1 GeV (\cite{lamb}). Meanwhile sources
  number 5 and 7 are 'Mrk 421' and '4C +10.45' respectively}
\end{table}

\end{document}